\documentclass{aa}
\usepackage{txfonts}
\usepackage{graphicx}
\usepackage{subcaption}
\usepackage{threeparttable} 
\usepackage[colorlinks=true,allcolors=blue]{hyperref}
\usepackage{gensymb}

\DeclareUnicodeCharacter{5415}{}  
\DeclareUnicodeCharacter{5EFA}{}  
\DeclareUnicodeCharacter{4F1F}{}  

\begin{document} 

\authorrunning{Mountrichas et al.}
\titlerunning{AGN obscuration: Nuclear vs. galactic gas and dust}

\title{AGN obscuration in optical and X-rays: Host properties and the interplay of nuclear and galactic gas and dust in a combined SDSS–XMM sample}

\author{
G.~Mountrichas\inst{1}, 
F.~J.~Carrera\inst{1}, 
E.~Quintin\inst{2}, 
A.~Viitanen\inst{3,4,5}, 
A.~Corral\inst{1}, 
N.~Webb\inst{6}
}

\institute{
\inst{1} Instituto de Física de Cantabria (CSIC–Universidad de Cantabria), Avenida de los Castros, 39005 Santander, Spain\\
\email{gmountrichas@gmail.com}\\
\inst{2} European Space Agency (ESA), European Space Astronomy Centre (ESAC), Camino Bajo del Castillo s/n,
28692 Villanueva de la Cañada, Madrid, Spain\\
\inst{3} Department of Astronomy, University of Geneva, ch.\ d’Ecogia 16, 1290 Versoix, Switzerland\\
\inst{4} INAF–Osservatorio Astronomico di Roma, Via Frascati 33, 00078 Monteporzio Catone, Italy\\
\inst{5} Department of Physics, Gustaf Hällströmin katu 2, 00014 University of Helsinki, Finland\\
\inst{6}  IRAP, Université de Toulouse, CNRS, UPS, CNES, 9 Avenue du Colonel Roche, BP 44346, 31028 Toulouse Cedex 4, France
}

\abstract{We investigate the link between optical obscuration and X-ray absorption in active galactic nuclei (AGN) by combining X-ray spectroscopy from 4XMM-DR11 with SDSS DR16Q spectroscopy. Bayesian X-ray spectral fits were obtained within the XMM2Athena project, and host-galaxy properties were derived via \textsc{CIGALE} SED fitting. Our final sample comprises 241 X-ray AGN at $z<1.9$. For 172 sources ($\sim70\%$), the optical broad-line (BL) or narrow-line (NL) classification agrees with their X-ray obscuration based on $N_{\rm H}$, but two mismatched populations emerge. Eleven BL AGN show signs of X-ray absorption (BLAbs) and elevated gas-to-dust ratios compared to BL AGN, consistent with dust-free or host-scale absorbers. Conversely, 58 NL AGN appear unobscured in X-rays (NLUnabs) and low gas-to-dust ratios. Nearly half are assigned type~1 properties by SED fitting, suggesting diluted or intrinsically weak broad-line regions, host contamination, or variability. Optical line diagnostics support this picture: NL AGN show higher Balmer decrements than NLUnabs, indicating stronger extinction and different ionization conditions. Host diagnostics further reinforce the contrasts: at $\rm z<0.8$, NLUnabs show star-formation rates and accretion efficiencies that are comparable to BL AGN, whereas NL AGN reside in more quiescent hosts with lower star formation and less efficient black-hole growth. BLAbs match BL AGN in host and accretion properties, with their peculiarity lying in excess X-ray absorption. These findings demonstrate that obscuration arises not only from orientation but also from multi-scale distributions of gas and dust. Identifying such mismatched populations will be crucial for interpreting AGN demographics in ongoing and upcoming surveys such as \emph{Euclid} and VRO/LSST.
}

   
\maketitle  

\section{Introduction}

Active galactic nuclei (AGN) are among the most powerful manifestations of accretion onto supermassive black holes (SMBHs) and play a fundamental role in the evolution of galaxies. Their energetic output, spanning from X-rays to the infrared (IR) and radio, can regulate the growth of galaxies by heating, expelling, or redistributing gas through feedback processes \citep[e.g.][]{Kormendy2013, Harrison2017}. AGN are observed across a wide range of luminosities and host environments, with diverse observational signatures in the X-ray, optical, and IR regimes \citep[e.g.][]{Urry1995, Hickox2018}. This diversity reflects the complex interplay between SMBH fueling, host-galaxy conditions, and obscuration by gas and dust.

Understanding the nature of AGN obscuration is therefore central to disentangling the connection between SMBHs and their hosts. In the simplest unification scenario, the observed AGN type depends primarily on orientation with respect to a circumnuclear dusty torus \citep{Antonucci1993, Urry1995}. However, it has become increasingly clear that orientation alone cannot explain the full range of observed obscuration properties. Recent models highlight the importance of additional ingredients, including evolutionary stages linked to gas inflows and outflows \citep[e.g.][]{Hopkins2006, Hickox2009, Ricci2017}, as well as the contribution of host-galaxy scale material such as dust lanes or circumnuclear star formation \citep[e.g.][]{Buchner2017, Malizia2020}. These “intermediate” scenarios posit that both nuclear and host-scale processes contribute to shaping obscuration.

A direct way to probe obscuration is to compare diagnostics at different wavelengths. X-rays trace line-of-sight gas through the equivalent neutral hydrogen column density ($N_H$), while optical and IR indicators are sensitive to dust extinction. Although several studies have found broad correlations between optical/IR obscuration and X-ray absorption \citep[e.g.][]{Civano2012, Merloni2014}, the relation shows substantial scatter \citep[e.g.][]{Jaffarian2020}. Multiple factors likely contribute, including: (i) variability in $N_{\rm H}$ on timescales of months to years \citep[e.g.][]{Reichert1985, Yang2016}, (ii) absorbing material on galactic scales unrelated to the nuclear torus \citep[e.g.][]{Goulding2012, Malizia2020}, (iii) differences in the column densities probed by X-ray, optical, and IR diagnostics \citep[e.g.][]{Masoura2020, Mountrichas2020}, and (iv) dust-to-gas ratios that deviate from the Galactic average, either due to dust-poor gas or dust-rich environments \citep[e.g.][]{Maiolino2001, Burtscher2016, Ricci2017}. As a result, significant populations of AGN are found to be obscured in one wavelength regime but not in another. Examples include optical type~1 AGN with strong X-ray absorption and, conversely, optically narrow-line AGN with little or no X-ray obscuration \citep[e.g.][]{Merloni2014, Lansbury2015, Mountrichas2020, Mountrichas2021b, BarquinGonzalez2024}.

These mismatches manifest in various forms: some AGN show strong X-ray absorption with little optical reddening (dust-free gas), others appear heavily X-ray absorbed yet still display broad UV/optical lines \citep[e.g.][]{Li2019}, and conversely, optically reddened AGN with absorbed SEDs may exhibit little or no X-ray obscuration \citep[e.g.][]{Masoura2020}. In some cases, the X-ray absorption exceeds that expected from the measured optical extinction \citep[e.g.][]{Granato1997, Merloni2014}. This complexity highlights the need for joint X-ray and optical analyses to disentangle nuclear from host-scale effects.

Another compelling piece of evidence for the complexity of AGN obscuration comes from the class of so-called Changing-Look AGN (CLAGN). These are systems that exhibit dramatic transitions between optical type~1 and type~2 classifications on timescales of months to years \citep[e.g.][]{LaMassa2015a, MacLeod2016, Yang2018, Ricci2020, Hon2022}. Such transitions are often driven by intrinsic changes in the accretion rate or structure of the broad-line region rather than by variable line-of-sight obscuration alone. Interestingly, X-ray variability in these sources does not always mirror the optical changes, suggesting that the mechanisms regulating obscuration and ionizing output may operate on distinct spatial or physical scales. The existence of CLAGN thus reinforces the notion that both orientation and intrinsic accretion variability contribute to the observed diversity of AGN classifications, complementing the population-wide approach adopted in this study.

The goal of this work is to investigate AGN populations that exhibit inconsistent X-ray and optical classifications, probing the physical origins of their obscuration using extensive X-ray spectral and optical spectroscopic data from wide-area surveys and host-galaxy properties derived via spectral energy distribution (SED) fitting. Our focus is on quantifying the agreement and disagreement between optical and X-ray classifications, characterizing the outlier populations, and exploring their gas-to-dust ratios, emission-line diagnostics, and host-galaxy properties. 

The paper is structured as follows. Section~\ref{sec_data} describes the data. Section~\ref{sec_analysis} outlines the sample selection and the classification methods. Section~\ref{sec_results} presents the main results and discusses the physical interpretation of AGN populations in the context of multi-scale obscuration. Section~\ref{sec_summary} summarizes our main findings. 

\section{Data}
\label{sec_data}

In this work we use X-ray AGN from the XMM–Newton 4XMM-DR11 catalogue \citep{Webb2020}, analyzed within the framework of the XMM2Athena project \citep{Webb2023, Viitanen2025}. The XMM2Athena pipeline provides homogeneous source classification, photometric redshifts, and X-ray spectral fitting for 4XMM sources, facilitating consistent multiwavelength analyses. The 4XMM-DR11 catalogue is the fourth generation of serendipitous X-ray source catalogues produced by the Survey Science Centre, containing $319,565$ detections from $11,907$ observations, corresponding to $210,444$ unique sources with spectra. Source classification follows the scheme of \citet{Tranin2022}, who cross-matched XMM detections with external catalogues of AGN, stars, and Galactic sources. Cross-matching the 4XMM-DR11 sources with their catalogue yielded $92,238$ classified objects, including $76,610$ AGN. Photometric redshifts were derived using the methodology of \citet{Ruiz2018}, which combines optical (SDSS or PanSTARRS) counterparts with near- and mid-IR data when available, employing the MLZ–TPZ machine-learning algorithm \citep{Kind2013}. Spectroscopic redshifts from SDSS are available for about $8,500$ sources, leading to $35,538$ AGN with redshift information overall \citep[see also][]{Viitanen2025}.

The X-ray spectral analysis was also carried out within the XMM2Athena framework using the BXA package \citep{Buchner2014}, which connects XSPEC \citep{Arnaud1996} with the nested-sampling algorithm \textsc{UltraNest} \citep{Buchner2019,Buchner2021}. The spectra were analysed in a Bayesian framework using uninformative priors, and the fits were performed with the Cash statistic, which is appropriate for Poisson-distributed data in the low-count regime. A redshifted absorbed power-law model with Galactic absorption, \texttt{cflux*tbabs(ztbabs*zpowerlaw)}, was applied. The Galactic hydrogen column density was fixed to the total column density along the line of sight, while the intrinsic absorber at the source redshift was left free. Background spectra were first modelled separately using the \texttt{bxa.sherpa.background} module, which accounts for instrumental, cosmic, and Galactic components, and the resulting background shape was then kept fixed during the source fitting apart from its normalisation. Mode values of the posterior distributions of $N_{\rm H}$, $\Gamma$, and fluxes (in the 2--10\,keV band) were adopted as the most probable parameter estimates, and intrinsic luminosities $L_{\rm X}$ were computed after correcting for absorption. Reliable X-ray spectral measurements are available for $30\,653$ AGN. Further details of the fitting procedure are given in \citet{Viitanen2025}.

The aim of our study is to compare X-ray absorption with optical obscuration and identify AGN populations that are optically obscured but X-ray unobscured, and vice versa. To this end, we use the \citet{Wu2022} catalogue, which provides emission-line and continuum properties for more than 750,000 quasars from SDSS DR16. 

The catalogue is based on a homogeneous spectral fitting procedure that models both the continuum and the main emission-line complexes (e.g.\ H$\alpha$, H$\beta$, Mg\,{\sc ii}, C\,{\sc iv}). For the Balmer lines (H$\alpha$, H$\beta$), the fits include multiple Gaussian components: a narrow core and one or more broad components when statistically justified. For H$\alpha$, the neighboring [N\,{\sc ii}] and [S\,{\sc ii}] lines are simultaneously modeled, while for H$\beta$ the [O\,{\sc iii}] doublet is fitted alongside. Similarly, Mg\,{\sc ii} and C\,{\sc iv} are decomposed into narrow and broad components, with additional Gaussians where necessary to capture asymmetric wings.  

For each line, \citet{Wu2022} provide both total and broad-component measurements (the latter denoted by the suffix \_BR). The total profile includes all fitted Gaussian components, while the \_BR quantities correspond to the isolated broad component when present. For narrow lines such as [O III] $\lambda5007$, they also report a “core” measurement obtained after subtracting potential blue wings. These measurements, particularly the total Full Width at Half Maximum (FWHM) of H$\beta$ and Mg II, together with the total [O III] properties, form the basis of our spectroscopic classifications in this work.

The host galaxy properties were derived through spectral energy distribution (SED) fitting with the \textsc{CIGALE} code \citep[][]{Boquien2019, Yang2020, Yang2022}. In this work, we use the measurements presented in \citet{Mountrichas2024c}. In brief, the stellar component was modelled with a delayed star-formation history (SFH) of the form ${\rm SFR}(t)\propto t\,\exp(-t/\tau)$, complemented by a starburst episode represented as a constant 50 Myr period of ongoing star formation \citep{Malek2018, Buat2019}. Stellar emission was generated using the single stellar population templates of \citet{Bruzual2003} and    attenuated according to the law of \citet{Charlot2000}. Nebular emission was added using the templates of \citet{VillaVelez2021}, while dust emission heated by stars was described with the models of \citet{Dale2014}, excluding any AGN contribution. The AGN emission itself was modelled with the \textsc{SKIRTOR} templates \citep{Stalevski2012, Stalevski2016}. Finally, \textsc{CIGALE} also accounts for X-ray emission, and in the fitting procedure the intrinsic $L_{\rm X}$ in the 2–10 keV band was used. The parameter space explored during the SED fitting is given in Table~1 of \citet{Mountrichas2024c}.

\section{Analysis}
\label{sec_analysis}
In this section, we describe the procedure followed to build the final AGN catalogue used in our analysis, together with the criteria adopted for the optical classification of the sources. We also present the scheme used to classify AGN based on measurements derived from SED fitting.

\subsection{Selection criteria}
\label{sec_criteria}

From the initial sample of 35\,538 AGN in our catalogue, we selected those flagged as \texttt{flag = 0} in the XMM2Athena catalogue. A detailed description of the flag definitions is provided in \citet{Viitanen2025}. Briefly, sources with \texttt{flag = 0} correspond to cases where both the background and source spectral fits have a $p-$value greater than 0.01, the adopted threshold for acceptable fits. This criterion yielded 30\,653 AGN.  

To ensure robust estimates of the host-galaxy properties, we required reliable multi-wavelength photometry. Specifically, we considered only sources with available measurements in SDSS or Pan-STARRS, 2MASS, and WISE, covering the following bands: $u, g, r, i, z, y, J, H, K, W1, W2$, and $W4$ 
\citep[e.g.][]{Mountrichas2021b, Mountrichas2021c, Buat2021, Mountrichas2022a, Mountrichas2022b, Mountrichas2023a, Mountrichas2024a}. This requirement was satisfied by 2\,460 AGN, all of which also have $W3$ photometry available, although it was not explicitly required. For these sources, we performed SED fitting using the templates and parameter space described in Sect.~\ref{sec_data}.  

To restrict the analysis to the most reliable host-galaxy measurements, we excluded poorly fitted SEDs by requiring a reduced $\chi^2_{red} < 5$. This threshold has been commonly adopted in previous works \citep[e.g.,][]{Masoura2018, Koutoulidis2022, Mountrichas2023b, Mountrichas2023c}, based on visual inspection of the SED quality. Approximately 82\% of the sources satisfied this condition.  

To ensure reliable estimates of the key \textsc{CIGALE} parameters used in our analysis, star formation rate (SFR), stellar mass ($M_\star$), AGN polar dust ($E_{B-V,\mathrm{AGN}}$), and dust attenuation in the interstellar medium ($A_{V,\mathrm{ISM}}$), we applied additional quality cuts following previous works \citep[e.g.][]{Mountrichas2021c, Buat2021, Mountrichas2023d, Mountrichas2024d, Mountrichas2024e}. These cuts rely on comparing the best-fit values with the Bayesian estimates provided by \textsc{CIGALE}. Specifically, we required: $\frac{1}{5} \leq \frac{{\rm parameter}_{\rm best}}{{\rm parameter}_{\rm bayes}} \leq 5,$
where ${\rm parameter}_{\rm best}$ denotes the best-fit value and ${\rm parameter}_{\rm bayes}$ the corresponding Bayesian estimate. Approximately 84\% of the sources satisfied these conditions, yielding a final sample of 1,694 AGN. We verified that modest variations in these limits, as explored in Mountrichas et al.\ (2021c), do not impact the results: changing the lower and upper bounds of the bayes over best ratio within the ranges $0.1$--$0.33$ and $3$--$10$ alters the final sample size by less than 5\%, confirming that our conclusions are robust against such adjustments.  

Finally, we restricted the analysis to sources at $\rm z < 1.9$. This redshift cut ensures that the optical emission lines used in our classification (e.g.\ H$\beta$, H$\alpha$, and Mg~II) remain accessible within the SDSS spectral coverage, while also providing sufficient numbers of sources in each bin for statistical analysis (see Sect.~\ref{sec_classificaiton}). The final sample contains 1\,591 X-ray AGN, of which 260 are also included in the \citet{Wu2022} catalogue.

\subsection{Optical classification}
\label{sec_classificaiton}

\begin{figure}
\centering
  \includegraphics[width=0.7\columnwidth, height=5.8cm]{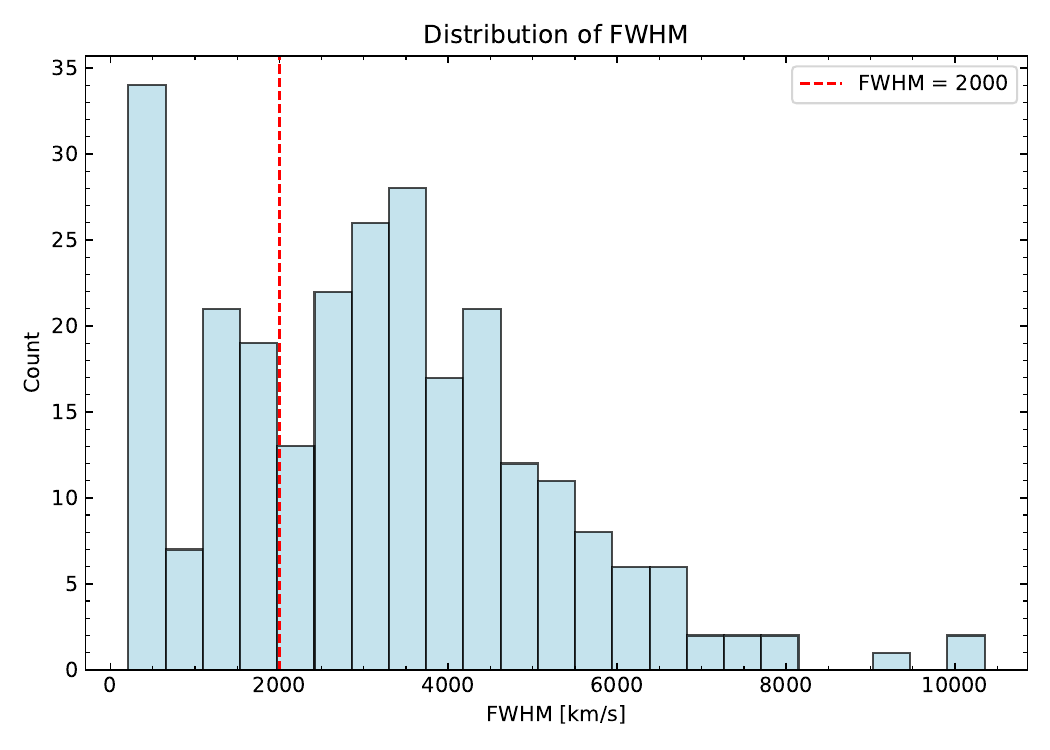}   
  \caption{Distribution of the FWHM for the 241 sources included in the analysis. The plotted value corresponds to the emission line used for each source (e.g., H$\beta$, Mg II), depending on redshift (see text for more details).}
  \label{fig_fwhm_distr}
\end{figure} 

\begin{figure}
\centering
  \includegraphics[width=1.\columnwidth, height=8.7cm]{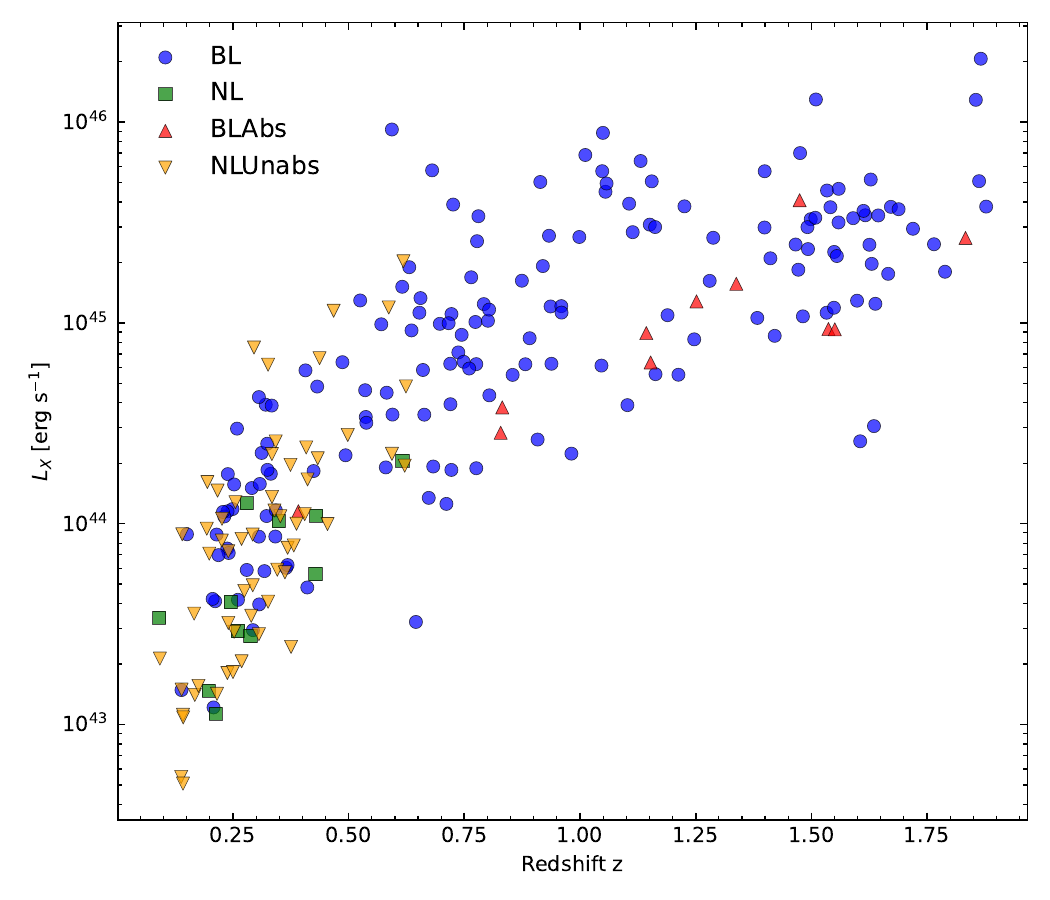}   
  \caption{Distribution of the 241 sources in the $L_{\mathrm{X}}$--$z$ plane. The definitions of Absorbed Broad (BLAbs) and Unabsorbed Narrow (NLUnabs) are given in Sect.~\ref{sec_optical_vs_xray}.}
  \label{fig_Lx_z}
\end{figure} 

\begin{figure}
\centering
  \includegraphics[width=0.8\columnwidth, height=6.cm]{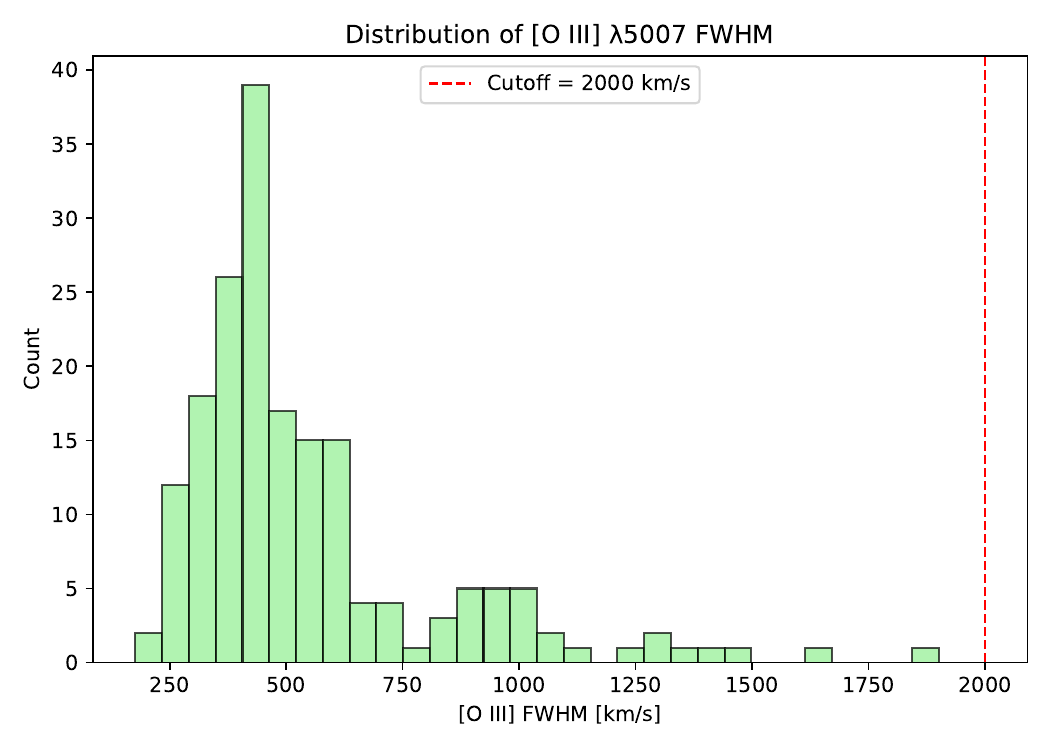}   
  \caption{Distribution of [O\,III]$\lambda5007$ FWHM values for the AGN sample. }
  \label{fig_OIII}
\end{figure} 

We first classify our X-ray selected AGN into broad-line (BL) and narrow-line (NL) sources, using the FWHM values provided by the \citet{Wu2022} catalogue. The choice of line depends on redshift: for sources at $\rm z<0.9$ we adopt the FWHM of H$\beta$, while for $\rm 0.9<z\leq 1.9$ we use Mg\,{\sc ii}. In addition to requiring that a value was reported, we applied two quality criteria: (i) the fractional FWHM error must be less than 50\% ($\sigma_{\rm FWHM}/{\rm FWHM}<0.5$), and (ii) the line detection significance must exceed ${\rm flux}/\sigma_{\rm flux}>2$. These cuts ensure that only robust measurements enter our classification. Among the 260 AGN in our parent sample, 241 have reliable FWHM measurements for these lines. 

For the classification, we adopt a threshold of $\rm 2000\,km\,s^{-1}$, a commonly used value in the literature \citep[e.g.][]{Shen2011, Lusso2012}, although somewhat different choices have also been applied \citep[e.g.][]{Menzel2016}. Fig. ~\ref{fig_fwhm_distr} shows the FWHM distribution of our sample, which exhibits a bimodal shape with a separation close to $\rm 2000\,km\,s^{-1}$, supporting the suitability of our adopted criterion. The narrow peak at very low FWHM values likely reflects the spectral resolution and the presence of sources with unresolved line widths. Among the 241 X-ray AGN, 172 are classified as BL and 69 as NL sources. Their distribution in the $L_{\mathrm{X}}$–$z$ plane is presented in Fig.~\ref{fig_Lx_z}. We note that most BL sources lie at higher redshift ($\rm z>1$), whereas NL AGN are scarce above $\rm z>0.7$.

In addition to the total FWHM, we inspected the individual line components fitted in the \citet{Wu2022} catalogue. Multi–Gaussian models were used for all lines, and a broad component was often included even when the physical presence of a BLR was uncertain. In most sources the broad and narrow components were consistent with the total line width, but a small number of cases showed a very broad component (FWHM $\gtrsim 5000$ km s$^{-1}$) while the total line width remained narrow ($<2000$ km s$^{-1}$). We interpreted such fits as artificial broad wings arising from the decomposition rather than genuine broad emission. These objects were classified as NL.

Conversely, if the total FWHM exceeded $2000\,{\rm km\,s^{-1}}$
, the source was classified as BL irrespective of the behaviour of the broad component. This criterion avoids misclassifications driven by spurious fits and ensures that our BL identification reflects genuine broad emission lines rather than artifacts of the fitting procedure. It also explains part of the difference with catalogues that rely solely on the broad component for classification.

We implemented this second check purely to guard against pathological broad-component fits in otherwise narrow lines. We further find that this second criterion yields classifications fully consistent with the first criterion described earlier. All sources classified as BL or NL on the basis of the total-line FWHM (with the $2000\,{\rm km\,s^{-1}}$ threshold and associated uncertainties) are recovered as such under the second criterion. This demonstrates that the two approaches are mutually reinforcing: the simple width threshold and the consistency check between total and broad components  both identify the same BL and NL populations.  We therefore adopt the simple total-FWHM threshold for all subsequent analysis.

Since [O\,III] is intrinsically a narrow emission line, its measured FWHM provides a natural benchmark against which to distinguish between broad- and narrow-line AGN. We therefore examined the distribution of [O\,III] FWHM in our sample (Fig. \ref{fig_OIII}). As expected, the values cluster at low widths ($\sim400$\,km\,s$^{-1}$), with the distribution extending up to $\sim2000$\,km\,s$^{-1}$, which corresponds to the maximum width observed for this narrow line. Any source whose Balmer or Mg\,II line exhibits an FWHM significantly above this upper envelope can therefore be reliably classified as a broad-line AGN, while values below this limit correspond to narrow-line systems. This empirical boundary is fully consistent with the threshold of 2000\,km\,s$^{-1}$ adopted in our classification based on the Balmer and Mg\,II lines, thereby reinforcing the robustness of our approach.  

\subsection{SED fitting classification}
\label{sec_cigale_class}

\cite{Mountrichas2021b} investigated X-ray AGN in the XMM-XXL field \citep{Pierre2016} and demonstrated that CIGALE can provide a reliable separation between type~1 and type~2 sources. Their classification was based on the best-fit and Bayesian estimates of the inclination angle $i$ derived by CIGALE: type~1 AGN correspond to $i_{\rm best}=30^{\circ}$ and $i_{\rm bayes}<40^{\circ}$, while type~2 AGN are identified with $i_{\rm best}=70^{\circ}$ and $i_{\rm bayes}>60^{\circ}$. They further showed that including polar dust in the SED fitting improves the reliability of type~2 identification, but also complicates the classification. In particular, a substantial fraction ($\sim$35\%) of AGN that appear as type~1 from their inclination angles, but have high polar dust extinction ($E_{BV,AGN}>0.15$), are spectroscopically classified as type~2. Motivated by these findings, subsequent studies have incorporated an $E_{BV,AGN}$ threshold in addition to inclination angle measurements when classifying AGN with CIGALE \citep[][]{Mountrichas2024a, Mountrichas2024b}. 

We note, however, that for the classification comparison in \citet{Mountrichas2021b}, they used the catalogue of \citet{Menzel2016}, in which a threshold of FWHM $>1000\,\rm km\,s^{-1}$ was adopted for broad-line identification, whereas in this work we apply a higher threshold (see Sect.~\ref{sec_classificaiton}). In the next section, we therefore redefine the SED-based classification criteria, introducing a new $E_{B-V,\mathrm{AGN}}$ threshold and inclination-angle boundary tailored to our sample, which we then use when comparing with the spectroscopic classifications.

\section{Results and discussion}
\label{sec_results}

In this section, we explore the relationship between optical obscuration and X-ray absorption, examine how these classifications compare with those derived from SED fitting based on inclination angle, and consider the relative roles of nuclear and host-galaxy gas and dust in shaping the observed properties.

\begin{figure}
\centering
  \includegraphics[width=0.9\columnwidth, height=8.cm]{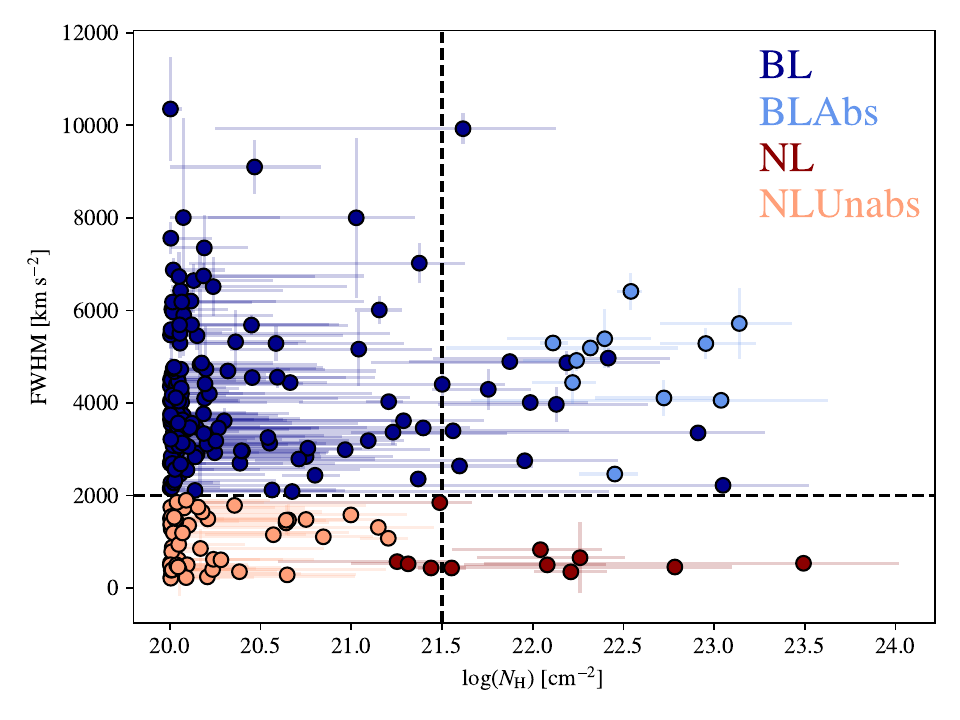}   
  \caption{FWHM versus X-ray absorption (N$\rm _H$). Different AGN populations are plotted using the colours and symbols indicated in the legend. The horizontal dashed line marks the FWHM threshold used to separate sources by the width of their optical lines, while the vertical dashed line indicates the N$\rm _H$ threshold adopted to identify X-ray–absorbed sources (see text for details).}
  \label{fig_fwhm_nh}
\end{figure}

\subsection{Optical and X-ray obscuration}
\label{sec_optical_vs_xray}

To investigate the connection between optical and X-ray obscuration in our dataset, we use the $N_{\rm H}$ parameter derived from Bayesian X-ray spectral fits and compare it with the optical classifications based on emission-line widths. Fig. ~\ref{fig_fwhm_nh} presents the distribution of sources in the FWHM–$N_{\rm H}$ plane.  Within this distribution, we identify two groups of outliers: (i) BL AGN that nonetheless show evidence of significant X-ray absorption, and (ii) NL AGN that appear virtually unabsorbed. We define the first class as BL AGN whose lower N$\rm _H$ bound exceeds $10^{21.5}\,$cm$^{-2}$ (i.e., N$\rm _H - N\rm _{H,err} > 10^{21.5}\rm{cm^{-2}}$; hereafter BLAbs). The second class includes NL AGN whose upper N$\rm _H$ bound falls below $10^{21.5}\,$cm$^{-2}$ (i.e., N$\rm _H + N\rm _{H,err} < 10^{21.5} \rm{cm^{-2}}$; hereafter NLUnabs). We note that, throughout the following analysis, the BL category refers exclusively to broad-line AGN without X-ray absorption, since BLAbs have been excluded from this group. An analogous separation is applied for NL AGN, with NLUnabs treated as a distinct population.
The choice of $N_{\rm H}=10^{21.5}\,{\rm cm}^{-2}$ to separate absorbed and unabsorbed sources is motivated by previous works showing that this limit provides a better correspondence with optical obscuration than the more conventional $10^{22}\,{\rm cm}^{-2}$ threshold  \citep[e.g.][]{Merloni2014} and has been also adopted in other studies \citep[e.g.,][]{Masoura2020, Masoura2021, Mountrichas2021c}. Therefore, adopting this boundary allows a more consistent comparison between X-ray and optical classifications, while remaining consistent with thresholds used in recent studies of AGN obscuration.

\begin{table}[ht]
\centering
\caption{Number of sources, median redshift and X-ray luminosity ($\log L_X$) for the AGN populations in different redshift bins.}
\label{table_class_redz}
\begin{tabular}{lccc}
\hline\hline
Population & $\rm N$ & $\rm z_{\rm med}$ & $\log L_{X,\,{\rm med}}$ [erg s$^{-1}$] \\
\hline
\multicolumn{4}{c}{$\rm z<0.4$} \\
Broad             & 36 & 0.31 & 44.05 \\
Narrow            & 8 & 0.27 & 43.57 \\
Absorbed Broad    &  1 & 0.39 & 44.06 \\
Unabsorbed Narrow & 45 & 0.29 & 43.89 \\
\hline
\multicolumn{4}{c}{$\rm 0.4 \leq z < 0.8$} \\
Broad             & 45 & 0.71 & 44.81 \\
Narrow            &  3 & 0.62 & 44.31 \\
Absorbed Broad    &  0 & --   & --    \\
Unabsorbed Narrow &  13 & 0.62 & 44.89 \\
\hline
\multicolumn{4}{c}{$\rm 0.8 \leq z < 1.9$} \\
Broad             & 80 & 1.47 & 45.43 \\
Narrow            &  0 & --   & --    \\
Absorbed Broad    &  10 & 1.41 & 45.04 \\
Unabsorbed Narrow &  0 & --   & --    \\
\hline
\end{tabular}
\end{table}

According to these criteria, our sample contains 11 BLAbs and 58 NLUnabs AGN. Their distribution in the $L_{\rm X}$–$\rm z$ plane is presented in Fig. \ref{fig_Lx_z} (see also Table \ref{table_class_redz}). Although, as already mentioned, NLUnabs AGN have been identified in other studies, too, we note that the majority of our NL AGN fall into the NLUnabs category. This is partly a consequence of our selection. First, our X-ray spectra are limited to sources with $\gtrsim 50$ counts, which naturally favors unobscured systems since absorbed spectra are fainter and less likely to meet this threshold. Second, our quality criteria (flag = 0, $p$-value > 0.01) tend to exclude strongly absorbed spectra, which are often noisy and more difficult to fit reliably. 
These effects bias the X-ray sample against absorbed AGN, leading to a dominance of NLUnabs sources.

To account for evolutionary effects, we analyze our subsets within distinct redshift bins. The binning scheme was determined objectively using the Bayesian blocks algorithm \citep{Scargle1998,Scargle2013}, which adaptively identifies change points in a distribution where the underlying event rate shifts, yielding edges near $\rm z\simeq0.40$, $0.80$, and $1.90$. These three bins provide a more uniform sampling of cosmic time than a simple two-bin split at $\rm z=0.8$. Adopting the \citet{Planck2018} cosmology, they correspond to lookback times of approximately 4.5~Gyr (0.0–0.4), 2.6~Gyr (0.4–0.8), and 3.1~Gyr (0.8–1.9), compared to 7.1 and 3.1~Gyr for a two-bin scheme. This configuration balances statistical power with a more even coverage of cosmic time, ensuring that comparisons between bins are not driven by systematic differences in $L_{\mathrm{X}}$ or $\rm z$.

We find that roughly half of the BL AGN lie at high redshift ($\rm z>0.8$), whereas almost all BLAbs (10 out of 11; $\sim$90\%) are found in this range (Table~\ref{table_class_redz}). In contrast, all NLUnabs are confined to $z<0.8$, with most (45 out of 58; $\sim$80\%) at $\rm z\leq0.4$. This difference largely reflects selection effects: BL AGN, with their strong unobscured continua and broad emission lines, are more easily detected and spectroscopically classified at higher redshift than NL AGN, whose host-dominated spectra limit their identification beyond $\rm z\sim0.8$. Nonetheless, the contrasting redshift distributions also hint that the two outlier populations may be influenced by distinct physical and observational factors.

Beyond these observational biases, the predominance of BLAbs at $\rm z>0.8$ could be linked to the higher gas fractions and more disturbed environments of galaxies at earlier epochs, where abundant gas reservoirs may obscure the X-rays while still allowing broad optical lines to be observed. Alternatively, these sources may represent AGN with complex absorber geometries or variable line-of-sight column densities \citep[e.g.][]{Reichert1985, Risaliti2002}, or cases where absorption arises on galactic rather than nuclear scales, such as dusty star-forming regions or large-scale gas clouds obscuring the X-ray emission \citep[e.g.][]{Goulding2011, Buchner2017}.

Conversely, the restriction of NLUnabs to lower redshifts may reflect obscuration dominated by dust rather than gas, resulting in elevated dust-to-gas ratios \citep[e.g.][]{Maiolino2001, Burtscher2016}. Another possibility is that these AGN host intrinsically weak or absent broad-line regions \citep[“true type 2” scenarios][]{Elitzur2009}, a phenomenon that becomes more common at lower luminosities (and thus lower redshifts). Orientation and variability effects may also play a role, suppressing the BLR in the optical without producing significant X-ray absorption \citep[e.g.,][]{LopezNavas2023}.

\subsection{SED based classification}
\label{sec_cigale_classif}

We also investigated the AGN classifications provided by SED fitting, which independently assigns each source to type 1, type 2, or Intermediate/Unclassified categories.  We adopted the following criteria to classify AGN using the inclination angle estimates ($i_{\rm best}$, $i_{\rm bayes}$) and the polar dust extinction ($E_{BV,AGN}$) derived by \textsc{CIGALE}, based on the analysis presented in \cite{Mountrichas2021b}:

\begin{itemize}
    \item type 1: $i_{\rm best} = 30^{\circ}$ and $i_{\rm bayes} \leq 40^{\circ}$, with $E_{BV,AGN} < 0.15$.
    \item type 2: 
    \begin{enumerate}
        \item $i_{\rm best} = 70^{\circ}$ and $i_{\rm bayes} \geq 60^{\circ}$, or
        \item $i_{\rm best} = 30^{\circ}$ and $i_{\rm bayes} \leq 40^{\circ}$, with $E_{BV,AGN} > 0.15$.
    \end{enumerate}
    \item Intermediate/Unclassified: all other cases.
\end{itemize}

Table~\ref{table_cigale_classif} presents the distribution of SED classifications for the full sample (260 AGN) and for the four AGN populations defined earlier (241 AGN with reliable FWHM measurements) compared to the spectroscopic classification. Applying these criteria, we find that CIGALE successfully identifies most of the spectroscopically confirmed NL AGN ($8/11$), with the remainder falling into the Intermediate/Unclassified category. For BL AGN, however, the recovery fraction decreases to $\sim 58\%$ ($94/161$). This lower efficiency is not unexpected, since in the analysis of \citet{Mountrichas2021b} the spectroscopic catalogue of \citet{Menzel2016} was used, where the adopted FWHM threshold for AGN classification is $\rm 1000\,km\,s^{-1}$, a value lower than the one applied in our study. 

To improve the match, we explored alternative thresholds for $E_{BV,AGN}$ (0.20, 0.25, 0.30) to quantify the trade-off between completeness and reliability, following the definitions in \citet{Mountrichas2021b}. Completeness measures the fraction of spectroscopically confirmed type~1 (or type~2) AGN also identified as such by SED fitting, while reliability refers to the fraction of SED-classified type~1 (or type~2) AGN confirmed by spectroscopy. We found that $E_{BV,AGN}=0.25$ provides the best overall compromise. In practice, this threshold substantially improves the recovery of BL AGN, while the type~2 recovery of NL AGN remains broadly similar within the uncertainties associated with the small sample size. Lowering the threshold to 0.20 increases completeness slightly but reduces reliability, while raising it to 0.30 has the reverse effect. This behaviour is consistent with earlier findings that including polar dust improves the recovery of obscured systems \citep{Mountrichas2021b}. We therefore adopt $E_{BV,AGN}=0.25$ as the optimal threshold in our refined classification scheme.

\begin{figure}
\centering
  \includegraphics[width=0.9\columnwidth, height=7.cm]{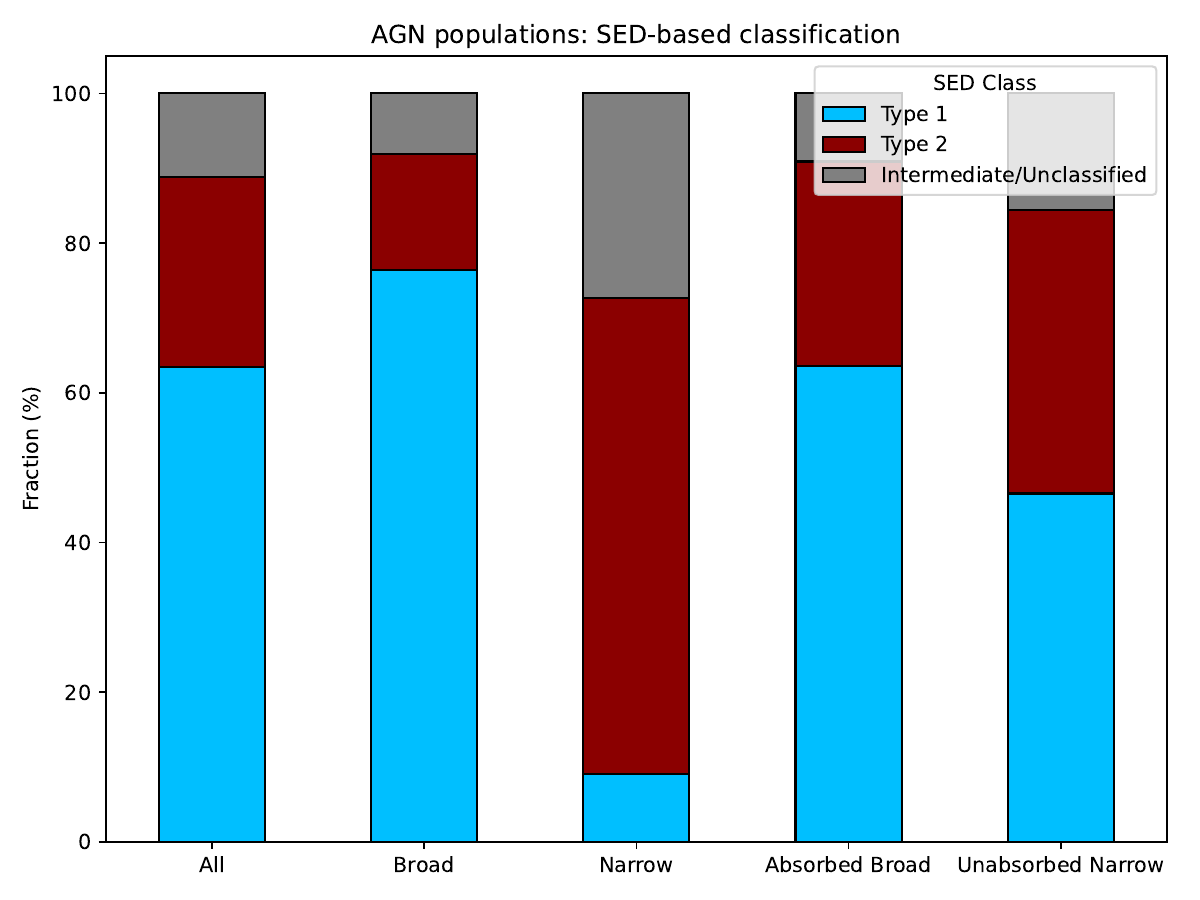}  
  \caption{Classification based on SED fitting for the four AGN populations, using the updated SED fitting criteria (see text for more details).}
  \label{fig_cigale_classif}
\end{figure}

The results of applying these updated criteria are shown in Table~\ref{table_cigale_classif_upd} and Fig.~\ref{fig_cigale_classif}. For the full sample, CIGALE assigns 63.5\% of sources to type~1, 25.3\% to type~2, and 11.2\% to Unclassified. Among the BL and NL populations, the main improvement relative to the original scheme is seen for BL AGN: 76.4\% are now identified as type~1, compared to 58.4\% for the $E_{BV,AGN}=0.15$ criterion. For NL AGN, the recovered type~2 fraction remains broadly similar, changing from 72.7\% to 63.6\%, that is, from 8/11 to 7/11 sources.

The analysis of the outlier populations provides further insight. Among BLAbs, seven out of eleven are classified as type~1 by CIGALE, consistent with unobscured SEDs and suggesting that their X-ray absorption arises from line-of-sight or host-scale gas rather than from material obscuring the BLR. Conversely, NLUnabs (NL AGN without X-ray absorption) are frequently assigned type~1–like properties ($\sim 47\%$), indicating SEDs dominated by unobscured emission despite the absence of broad optical lines. Possible explanations include host-galaxy dilution of weak broad lines, intrinsic BLR weakness (“true type~2” scenarios), or variability. 

In summary, the updated SED-based classification improves the agreement with the spectroscopic separation mainly for BL sources, while preserving a broadly similar performance for NL sources given the small numbers involved. At the same time, it provides an independent diagnostic that highlights genuine mismatched populations. These results demonstrate that SED fitting complements optical and X-ray methods, revealing cases where obscuration properties cannot be fully captured by spectroscopy alone.

\setlength{\tabcolsep}{2.pt} 
\begin{table}[ht]
\centering
\caption{Comparison between SED-based AGN classifications (using the criteria of \citealt{Mountrichas2021b}, i.e. $E_{BV,AGN}=0.15$) and spectroscopic classifications.}
\begin{tabular}{lccccc}
\hline\hline
Population & $N$ & type 1 & type 2 & Int./Unc. \\
\hline
All              & 260 & 113 (43.5\%) & 118 (45.3\%) & 29 (11.2\%) \\
BL               & 161 & 94 (58.4\%)  & 54 (33.5\%)  & 13 (8.1\%)  \\
NL               & 11  & 0 (0.0\%)    & 8 (72.7\%)   & 3 (27.3\%)  \\
BLAbs            & 11  & 3 (27.3\%)   & 7 (63.6\%)   & 1 (9.1\%)   \\
NLUnabs          & 58  & 14 (24.1\%)  & 35 (60.3\%)  & 9 (15.5\%)  \\
\hline
\label{table_cigale_classif}
\end{tabular}
\end{table}

\setlength{\tabcolsep}{2.pt} 
\begin{table}[ht]
\centering
\caption{Comparison between SED-based AGN classifications using the updated criteria ($E_{BV,AGN}$ threshold is set to 0.25, see text for more details) and spectroscopic classifications.}
\begin{tabular}{lccccc}
\hline\hline
Population & $N$ & type 1 & type 2 & Int./Unc. \\
\hline
All              & 260 & 165 (63.5\%) & 66 (25.3\%) & 29 (11.2\%) \\
BL               & 161 & 123 (76.4\%) & 25 (15.5\%) & 13 (8.1\%)  \\
NL               & 11  & 1 (9.1\%)    & 7 (63.6\%)  & 3 (27.3\%)  \\
BLAbs            & 11  & 7 (63.6\%)   & 3 (27.3\%)  & 1 (9.1\%)   \\
NLUnabs          & 58  & 27 (46.6\%)  & 22 (37.9\%) & 9 (15.5\%)  \\
\hline
\label{table_cigale_classif_upd}
\end{tabular}
\end{table}

\subsection{Gas-to-dust ratios}
\label{gas_dust_ratio}

\begin{figure}[ht]
\centering
\includegraphics[width=0.49\textwidth]{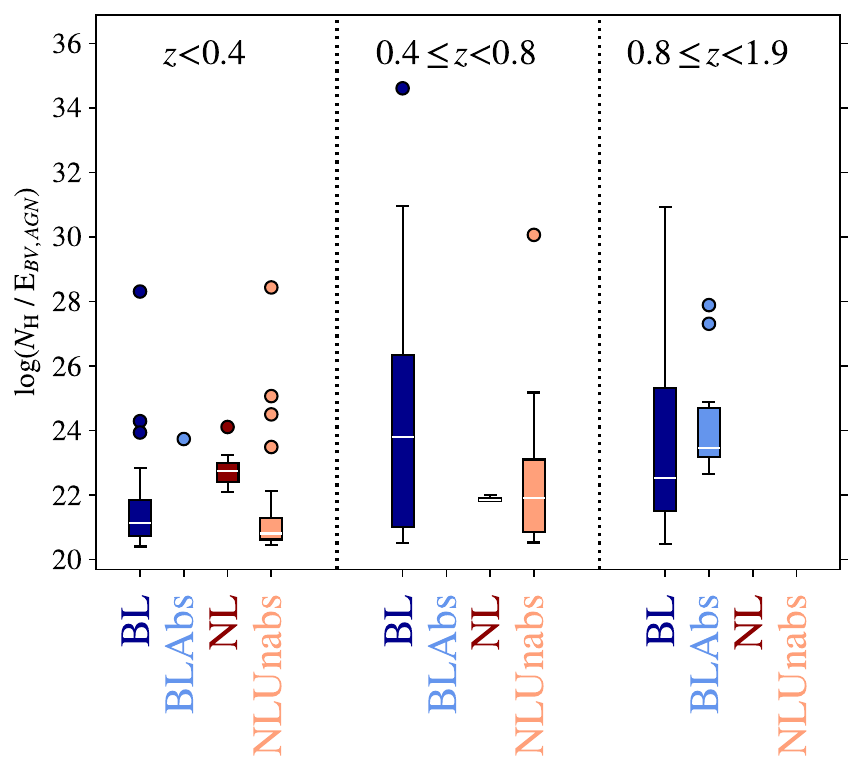}
\includegraphics[width=0.49\textwidth]{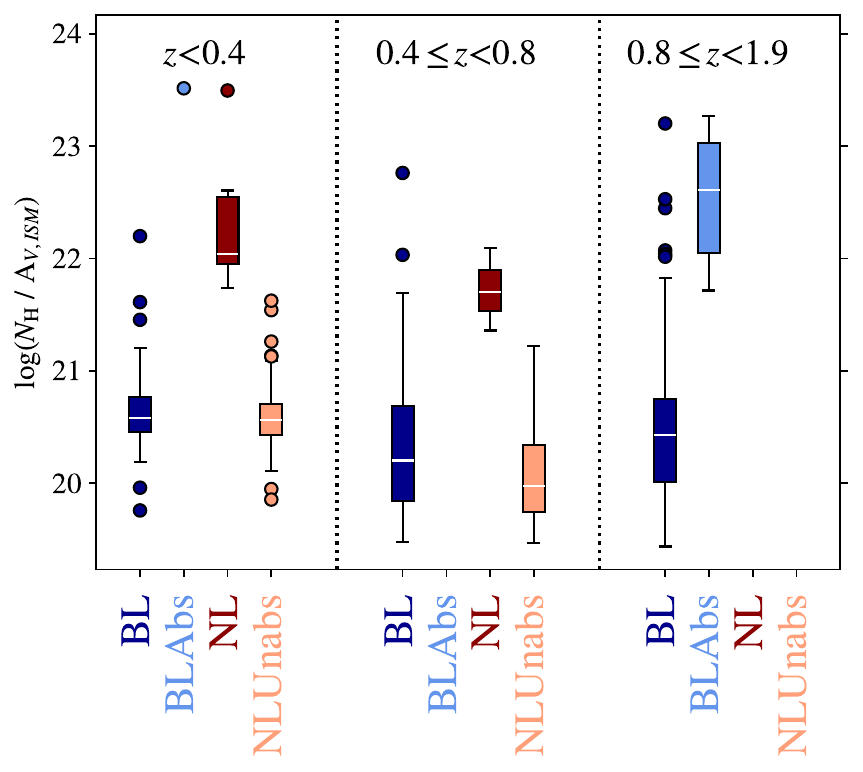}
\caption{Gas-to-dust ratios for the four AGN populations, expressed as $\log(N_{\rm H}/E(B\!-\!V)_{\rm AGN})$ (top panel) and $\log(N_{\rm H}/A_V^{\rm ISM})$ (bottom panel). Whiskers extend to $1.5\times\mathrm{IQR}$, and individual points beyond this range are shown as single circles.  
Populations with only one available measurement (e.g. BLAbs at $z<0.4$) are shown as single points to indicate the lack of statistical constraints. Median values and interquartile ranges for different redshift intervals are presented in Table \ref{tab:g2d_summary}.}
\label{fig_g2d}
\end{figure}

We used the \textsc{CIGALE} SED fitting results to examine gas-to-dust conditions through the ratios $N_{\rm H}/E_{BV, \mathrm{AGN}}$ and $N_{\rm H}/A_V^{\rm ISM}$, expressed as $\log(N_{\rm H}/A)$ with units of cm$^{-2}$\,mag$^{-1}$ (Fig.~\ref{fig_g2d} and Table~\ref{tab:g2d_summary}). These ratios are not dimensionless, but provide a relative indicator of the amount of gas per unit dust extinction, usually compared to the Galactic value of $N_{\rm H}/E_{(B{-}V)}\simeq5.8\times10^{21}$,cm$^{-2}$,mag$^{-1}$.

Because $E_{BV, {\rm AGN}}$ traces polar dust and is most meaningful for systems viewed close to face-on, we restrict the comparison of $N_{\rm H}/E_{BV, \mathrm{AGN}}$ to BL and BLAbs sources. At $0.8 \leq z < 1.9$, BLAbs show significantly higher values than BL AGN, with median $\log(N_{\rm H}/E_{(B{-}V),{\rm AGN}}) = 22.6$ compared to $20.4$ for BL. A Kolmogorov-Smirnov test (KS) yields $p \sim 5\times10^{-6}$. This confirms that BLAbs are genuinely gas-rich systems with little associated dust, consistent with dust-poor absorbers, clumpy geometries, or complex circumnuclear structure.

To compare all four populations we use $N_{\rm H}/A_V^{\rm ISM}$, which traces host-galaxy scale dust. At $z<0.4$, NL AGN show median $\log(N_{\rm H}/A_V^{\rm ISM}) = 22.0$ versus 20.6 for BL AGN ($p\sim 10^{-7}$), reflecting their higher $N_{\rm H}$ values (median $\log N_{\rm H}=22.1$ versus 20.0, see Table \ref{tab:nh_summary}). By contrast, NLUnabs show much lower values, driven by their small columns (median $\log N_{\rm H}=20.0$). Thus NLUnabs represent dust-rich but gas-poor systems, consistent with the absence of X-ray absorption despite optical type~2 spectra.

An intriguing possibility is that the elevated gas-to-dust ratios we observe, particularly for the BLAbs at $\rm z\gtrsim0.8$, may trace recent or ongoing interactions that funnel large columns of relatively dust-poor gas to the nucleus. Merger-driven inflows can (i) raise $N_{\rm H}$ faster than the dust content if the accreted gas is metal-poor or if dust is partially destroyed by merger-induced shocks and intense radiation fields, and (ii) produce clumpy, multi-scale geometries in which X-ray sightlines intercept dense gas while optical/UV continuum and broad lines remain only mildly extincted. In this context, high $N_{\rm H}/E_{BV, \mathrm{AGN}}$ may flag systems where the circumnuclear medium has been recently re-supplied or re-shaped by dynamical disturbances rather than by a steady-state torus.

Such conditions are also favourable for the presence of close SMBH pairs (dual AGN) at kpc scales. If some of our high gas-to-dust outliers are late-stage mergers, a dual–AGN interpretation is plausible in a subset of cases. However, our current datasets cannot test this directly: the XMM–Newton PSF and SDSS fiber spectroscopy do not resolve kpc-scale pairs at the redshifts of interest, and the [OIII]/H$\beta$ and Balmer-decrement diagnostics we use are not uniquely predictive of dual activity.

In summary, the gas-to-dust ratios derived from SED fitting confirm and sharpen the contrasts between BL, NL, and the outlier populations, while also hinting at possible evolutionary or dynamical origins. At low redshift ($z<0.4$), NL AGN display systematically higher $N_{\rm H}$ per unit dust extinction than BL AGN, largely reflecting their elevated gas columns. BLAbs at $z>0.8$ appear extremely gas-rich relative to their dust content, whereas NLUnabs at $z<0.8$ show the opposite behaviour, with low gas-to-dust ratios driven by their low $N_{\rm H}$. These findings indicate that the mismatched populations are not artifacts of classification but reflect genuine physical diversity in the circumnuclear and host-scale environments. The high gas-to-dust ratio systems may plausibly trace recent or ongoing mergers that channel relatively dust-poor gas toward the nucleus, raising the possibility that some could host dual AGN. Although our data cannot confirm this scenario, it remains testable with high-resolution X-ray, radio, or IFU observations capable of identifying double nuclei or disturbed kinematics.

\begin{table*}[ht]
\centering
\begin{threeparttable}
\caption{Median gas-to-dust ratios (with IQR in brackets) 
for the AGN populations in different redshift bins, and for the combined $\rm z<1.9$ sample.}
\label{tab:g2d_summary}
\begin{tabular}{lccc|ccc}
\hline\hline
 & \multicolumn{3}{c}{$\log(N_{\rm H}/E(B{-}V)_{AGN})$} & \multicolumn{3}{c}{$\log(N_{\rm H}/A_{V,ISM})$} \\
Population &  $\rm z<0.4$ & $\rm 0.4 \leq z < 0.8$ & $\rm 0.8 \leq z < 1.9$  & $\rm z<0.4$ & $\rm 0.4 \leq z < 0.8$ & $\rm 0.8 \leq z < 1.9$ \\
\hline
BL              & 21.1 [20.7, 21.9] & 22.8 [22.0, 24.3] & 22.5 [21.5, 25.3]  & 20.6 [20.5, 20.8] & 20.2 [19.8, 20.7] & 20.4 [20.0, 20.8] \\
NL             & 22.8 [22.4, 23.0] & 23.1 [22.8,23.9] & --  & 22.0 [21.9, 22.5] & 21.7 [21.5, 21.9] & -- \\
BLAbs     & --\tnote{a}      & -- & 23.5 [23.2, 24.7] &   --\tnote{a} & -- & 22.6 [22.0, 23.0] \\
NLUnabs  & 20.8 [20.6, 21.3] & 20.9 [20.1, 22.1] & -- & 20.6 [20.4, 20.7] & 20.0 [19.7, 20.3] & -- \\
\hline
\end{tabular}
\begin{tablenotes}
\footnotesize
\item[a] Values correspond to single-source bins and are not statistically meaningful and have been omitted. 
\item All gas-to-dust ratios are expressed as $\log(N_{\rm H}/A)$ in units of cm$^{-2}$\,mag$^{-1}$.
\end{tablenotes}
\end{threeparttable}
\end{table*}

\begin{table*}[ht]
\centering
\begin{threeparttable}
\caption{Median $\log N_{\rm H}$ values ($\mathrm{cm^{-2}}$; with IQR in brackets) for the AGN populations in different redshift bins.}
\label{tab:nh_summary}
\setlength{\tabcolsep}{2pt}
\begin{tabular}{lccc}
\hline\hline
Population & $\rm z<0.4$ & $0.4 \leq z < 0.8$ & $0.8 \leq z < 1.9$ \\
\hline
BL             & 20.0 [20.0,20.2] & 20.1 [20.0,20.3] & 20.1 [20.0,20.6] \\
NL             & 22.1 [21.9,22.4] & 21.3 [21.3,21.4] & -- \\
BLAbs        & --\tnote{a} & -- & 22.4 [22.2,22.7] \\
NLUnabs     & 20.1 [20.0,20.3] & 20.1 [20.0,20.1] & -- \\
\hline
\end{tabular}
\begin{tablenotes}
\footnotesize
\item[a] Single-source bin; not statistically meaningful.
\end{tablenotes}
\end{threeparttable}
\end{table*}

\subsection{Optical Line Diagnostics: Balmer Decrement and [O\,III]/H$\beta$ Ratios}

Optical emission-line flux ratios provide an independent probe of the dust and gas content along the line of sight, complementary to the gas-to-dust ratios discussed above. Here we focus on two diagnostics: the Balmer decrement, defined as the flux ratio H$\alpha$/H$\beta$, and the [O III]/H$\beta$ flux ratio. Both are sensitive to reddening, though in different ways. The Balmer decrement is computed using the narrow components of both lines, tracing dust attenuation in the ionized gas of the narrow-line region. In contrast, the [O III]/H$\beta$ ratio is computed from the narrow [O III]$_{\rm c}$ flux (as denoted in the catalogue) and the broad H$\beta$ flux and thus can only be evaluated for broad-line (type 1) AGN. In this context, it serves primarily as a qualitative indicator of differential extinction between the narrow- and broad-line regions rather than a classical ionization diagnostic.

A caveat is that these measurements rely on the fluxes provided in the catalogue, and not all spectra have reliable decompositions for the relevant emission lines. Consequently, the number of available sources is somewhat smaller than in the gas-to-dust ratio analysis (Table \ref{tab:counts_balmer_oiii}).

\begin{table}[ht]
\centering
\begin{threeparttable}
\caption{Counts of sources per AGN class and redshift interval for which either Balmer decrement or [O III]$_c$/H$\beta$ measurements are available.}
\label{tab:counts_balmer_oiii}
\setlength{\tabcolsep}{5pt}
\begin{tabular}{lcccc}
\hline\hline
Population & $\rm z{<}0.4$ & $\rm 0.4{\leq}z{<}0.8$ & $\rm 0.8{\leq}z{<}1.9$ & All \\
\hline
BL             & 36 & 45 & 17 & 98 \\
NL            & 8 &  3 &  0 & 11 \\
BLAbs    &  1 &  0 &  1 &  2 \\
NLUnabs & 45 &  13 &  0 & 58 \\
\hline
\end{tabular}
\end{threeparttable}
\end{table}

\subsubsection{Balmer decrement}

\begin{figure}[ht]
\centering
\includegraphics[width=0.49\textwidth]{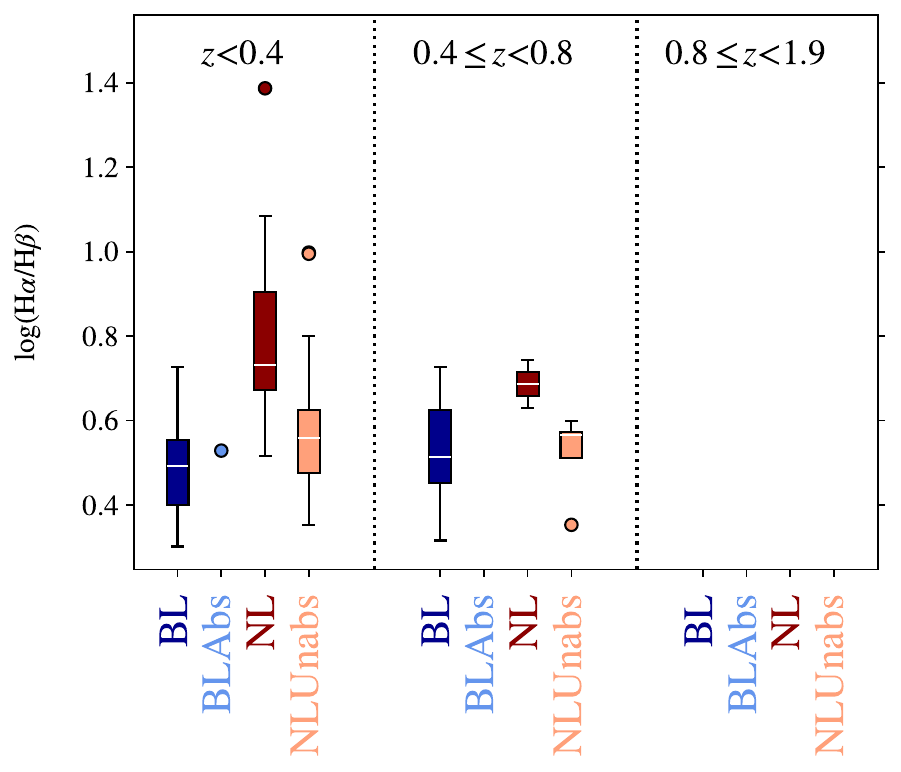}
\caption{Box plots of the Balmer decrement ($\mathrm{H\alpha/H\beta}$) for the BL, BLAbs, NL, 
and NLUnabs populations, shown in three redshift intervals. 
The boxes indicate the interquartile range (IQR), with the median marked by a horizontal line. 
Whiskers extend to $1.5\times\mathrm{IQR}$, and individual points beyond this range are shown as single circles. 
For subclasses represented by a single measurement (e.g. BLAbs at $z<0.4$), the data point is plotted directly.}
\label{fig_balmer}
\end{figure}

The Balmer decrement, H$\alpha$/H$\beta$, provides a direct probe of dust reddening in the narrow-line region (NLR). Our resuls are presented in Fig. \ref{fig_balmer}.  At $\rm z<0.4$, where sufficient numbers of NL AGN are available, NL sources show systematically higher Balmer decrements (median: $5.40$, IQR: [4.72, 8.29]) compared to NLUnabs (median: $3.62$, IQR: [2.99, 3.38]). This difference is statistically significant (KS test $p\simeq6\times10^{-3}$; $\sim2.7\sigma$), indicating moderately stronger dust extinction in the NL population. At higher redshifts, the small number of NL AGN with measurable H$\alpha$/H$\beta$ prevents meaningful statistical comparisons.

Physically, this result implies that NL AGN have more substantial columns of dusty material affecting the NLR emission. The excess extinction could arise from circumnuclear structures such as outflowing material or polar dust, or from host-scale dust lanes intersecting the line of sight. High-resolution imaging has shown that dust lanes and tidal features on kpc scales can indeed cover a significant fraction of the ionized gas region (e.g. \citealt{Goulding2012, Prieto2014}), and therefore can obscure the NLR when viewed along particular orientations. In contrast, NLUnabs appear to represent systems where the NLR is comparatively unobscured by dust.

\subsubsection{[O\,III]/H$\beta$ ratios}

Regarding the [O III]/H$\beta$ ratio, there are only two BLAbs sources in our sample that have available measurements, one at $\rm z<0.4$ and one at $\rm 0.8 \le z < 1.9$. In both redshift bins their [O III]/H$\beta$ ratios appear lower than those of the BL population, but the numbers are far too small for any statistical inference. If this tentative behaviour were confirmed in larger samples, it could indicate comparatively stronger broad H$\beta$ emission relative to the narrow-line component in BLAbs, consistent with their elevated gas-to-dust ratios and the presence of dust-poor, X-ray absorbing gas. However, with only two objects, this interpretation remains speculative.

\begin{figure}[ht]
\centering
\includegraphics[width=0.4\textwidth, height=5 cm]{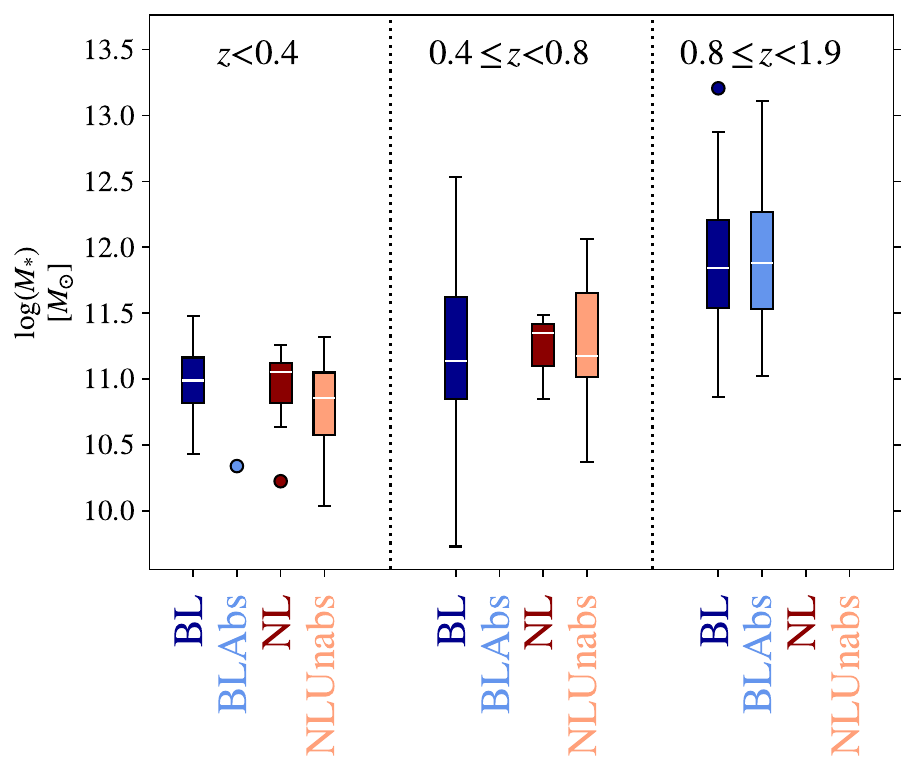}
\includegraphics[width=0.4\textwidth, height=5 cm]{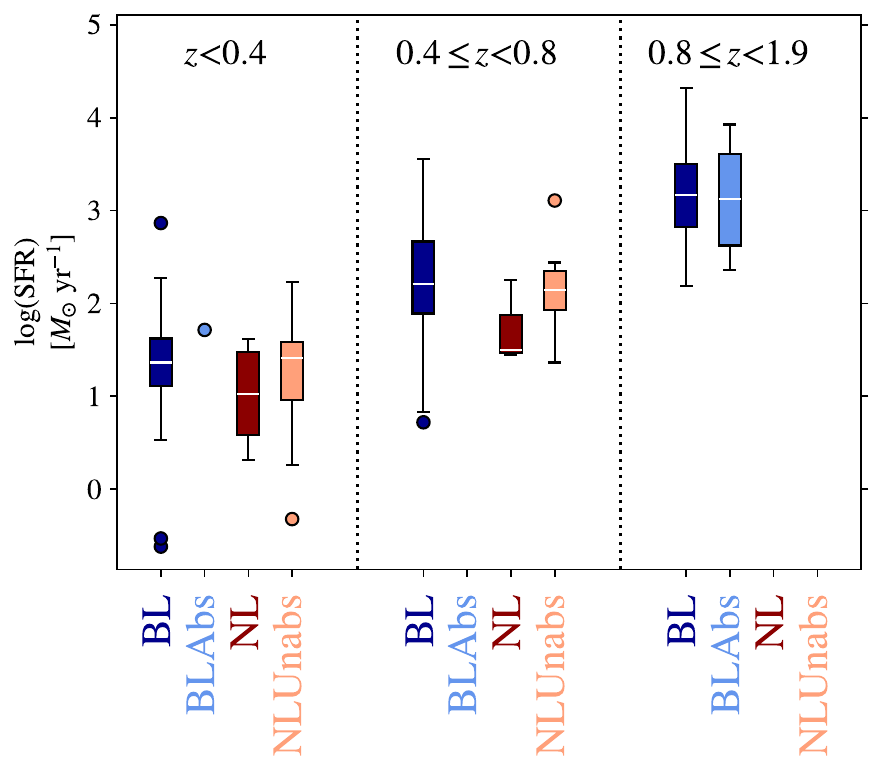}
\includegraphics[width=0.4\textwidth, height=5 cm]{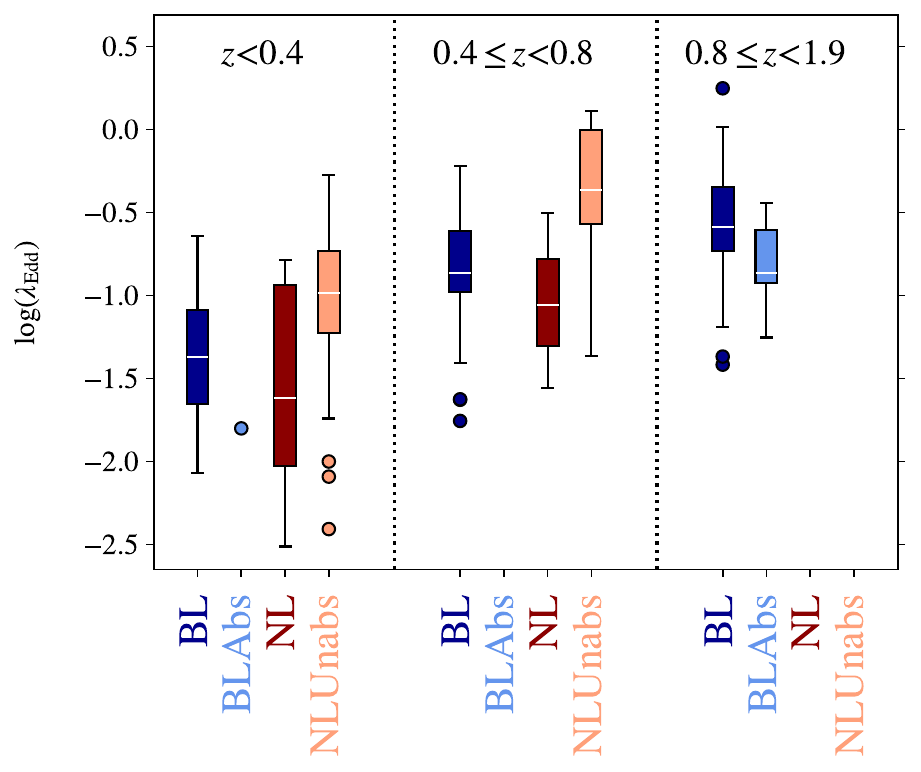}
\caption{Box plots of host galaxy stellar mass ($M_\star$, top), star formation rate (SFR, middle), and Eddington ratio ($\lambda_{\rm Edd}$, bottom) for the four AGN populations, in three redshift intervals. Whiskers extend to $1.5\times\mathrm{IQR}$, and individual points beyond this range are shown as single circles. For classes represented by a single measurement (e.g.\ BLAbs at low redshift), the individual data point is plotted directly. Median values and interquartile ranges for different redshift intervals are presented in Tables \ref{tab:host_mass}, \ref{tab:host_sfr} and \ref{tab:host_edd}.}
\label{fig_host_properties}
\end{figure}

\subsection{Host galaxy properties and accretion rates}

We then explore how $M_\star$, SFR, and Eddington ratio ($\lambda_{\mathrm{Edd}}$) vary across the AGN samples. The $M_\star$ and SFR estimates are obtained through SED fitting, and the $\lambda_{\mathrm{Edd}}$ values are adopted from the \citet{Wu2022} catalogue.

Fig. \ref{fig_host_properties} presents our measurements and Tables \ref{tab:host_mass}, \ref{tab:host_sfr}, and \ref{tab:host_edd} list the median and interquartile range (IQR) values in each redshift bin.
Black-hole masses were taken from the catalogue of \citet{Wu2022}, which provides virial $M_{\rm BH}$ estimates for quasars in SDSS DR16Q based on the widths of broad emission lines (H$\beta$, Mg II, C IV). We note that, although our spectroscopic classification identifies some of these sources as NL or NLUnabs, all objects in \citet{Wu2022} were originally classified as broad-line quasars. Their $M_{\rm BH}$ and $\lambda_{\rm Edd}$ values are therefore derived under the assumption of type~1 AGN physics. This does not affect the $M_\star$ and SFR measurements discussed below, but it does imply that the absolute $M_{\rm BH}$ and $\lambda_{\rm Edd}$ values of the NL and NLUnabs subsets are more uncertain than for the BL population. In particular, if the broad-line widths used in the virial estimators are underestimated in these borderline systems, the inferred $M_{\rm BH}$ may be biased low and the corresponding $\lambda_{\rm Edd}$ biased high. The Eddington-ratio trends for NL and NLUnabs should therefore be interpreted in a relative rather than absolute sense.

At $\rm z<0.8$, NL AGN appear to reside in somewhat more massive galaxies than BL AGN, although the offsets are modest ($\lesssim0.2$\,dex) and not statistically significant. NLUnabs have host masses comparable to the other two populations. These results align with earlier optical studies that reported small ($0.2$–$0.3$\,dex) mass differences between BL and NL AGN \citep[e.g.][]{Zou2019, Mountrichas2021b}, though often without strong statistical significance. X-ray–based classifications show similarly mixed results, with some analyses reporting negligible differences \citep[e.g.][]{Masoura2021, Mountrichas2021c} and others finding obscured sources in more massive hosts, particularly at moderate luminosities and high $N_{\rm H}$ \citep[e.g.][]{Georgantopoulos2023, Mountrichas2024c}.

SFRs reveal somewhat clearer contrasts. At both low ($\rm z<0.4$) and intermediate ($\rm 0.4 \leq z < 0.8$) redshifts, NL AGN exhibit systematically lower SFRs than BL AGN (median offsets of $\sim0.3$–0.7\,dex; KS test $p\simeq10^{-2}$, $\sim2.4\sigma$), indicating a mild but consistent difference. NLUnabs, in contrast, display SFRs more closely aligned with those of BL AGN: at $z<0.4$ their median SFR is 1.41 compared to 1.36 for BL, whereas NL AGN are lower at 1.03. A similar pattern holds at $\rm 0.4 \leq z < 0.8$, with NLUnabs (2.14) resembling BL (2.21) rather than NL (1.50). Although the differences do not exceed the 3$\sigma$ level, they suggest that NLUnabs share host star-forming properties with BL AGN rather than with the more quiescent NL population. At higher redshifts, BLAbs show SFRs comparable to those of BL AGN. Previous studies have likewise reported that unobscured AGN tend to exhibit similar SFRs to their obscured counterparts \citep[e.g.][]{Zou2019, Mountrichas2021b, Mountrichas2021c, Mountrichas2024c}, except in heavily absorbed, moderate-luminosity systems where suppressed star formation has been observed \citep[e.g.][]{Georgantopoulos2023}.

Eddington ratios show the most striking population differences and a strong redshift evolution. BL AGN increase from $\log \lambda_{\rm Edd}\sim -1.4$ at $\rm z<0.4$ to $\sim -0.6$ at $\rm z>0.8$, consistent with more efficient accretion at earlier cosmic times. NL AGN at $\rm z<0.4$ have lower accretion rates (median $-1.62$), pointing to less efficient black hole growth, although this difference is not statistically significant ($p\sim0.4$). NLUnabs show substantially higher Eddington ratios than both BL and NL at the same redshifts ($p \sim 10^{-3}$), reaching medians of $-0.99$ at $\rm z<0.4$ and $-0.37$ at $0.4 \leq z < 0.8$. This suggests a population of narrow-line systems undergoing relatively efficient accretion despite their dust-dominated signatures, although the absolute $\lambda_{\rm Edd}$ values for these subsets should be treated with caution because their virial $M_{\rm BH}$ estimates were derived under type~1 assumptions. BLAbs at $\rm z>0.8$ show Eddington ratios ($-0.86$) consistent with the BL population, suggesting that their distinctive property is the presence of excess X-ray–absorbing gas rather than differences in accretion efficiency.

The lower SFRs of NL AGN relative to BL sources are consistent with their lower Eddington ratios, pointing to more evolved hosts with reduced cold-gas reservoirs and less efficient black-hole fueling. This interpretation is consistent with their elevated gas-to-dust ratios and high Balmer decrements, which indicate substantial gas columns but relatively little dust, implying that obscuration and fueling are partially decoupled. By contrast, NLUnabs exhibit star-formation levels comparable to BL AGN, suggesting that they reside in actively star-forming hosts where the absence of broad lines is not linked to global quenching but rather to line-of-sight dust or host-dilution effects. BLAbs, which match the BL population in SFR and accretion rates at high redshift, reinforce the view that their X-ray absorption is driven by external or geometrically complex absorbers rather than by differences in fueling.

We also examined the long-term X-ray variability of the sources using the X-ray multi-mission catalogue from the XMM2Athena project \citep{Quintin2024}, which provides flux-based variability estimates across multiple epochs. No significant differences were found among the four AGN populations within any redshift bin, apart from a tentative indication that BLAbs at $\rm z>0.8$ appear less variable than their unabsorbed counterparts. This trend, however, is likely driven by lower signal-to-noise ratios in absorbed sources rather than by intrinsic variability differences. Overall, X-ray variability does not appear to be a distinguishing factor among the populations considered here.

Taken together, these host and accretion diagnostics reinforce the broader picture. NL AGN are dust- and gas-rich systems with relatively inefficient accretion at low redshift. NLUnabs combine dust-dominated signatures with apparently high Eddington ratios and BL-like star formation, marking them as a distinct population, although the absolute $\lambda_{\rm Edd}$ values for this subset remain more uncertain than for BL AGN. BL AGN show steadily increasing accretion efficiency with redshift, consistent with the canonical unobscured AGN sequence reported in previous works \citep[e.g.][]{Shen2012,Schulze2015,Aird2018}, while BLAbs highlight the presence of X-ray absorption unrelated to fueling differences. These complementary diagnostics suggest that the mismatched populations reflect genuine diversity in AGN fueling and obscuration rather than classification uncertainties. We note that these trends are unlikely to arise from selection effects: our comparison samples occupy similar $L_{\rm X}$ and redshift ranges within each bin, and the contrasts persist when luminosity-matched subsamples are considered, indicating that the differences are intrinsic.

\begin{table}[ht]
\centering
\begin{threeparttable}
\caption{Median stellar masses ($\log M_\star \,[M_\odot]
$) with interquartile ranges (IQR) for the AGN populations in different redshift bins.}
\label{tab:host_mass}
\setlength{\tabcolsep}{1pt}
\begin{tabular}{lccc}
\hline\hline
Population & $\rm z<0.4$ & $\rm 0.4 \leq z < 0.8$ & $\rm 0.8 \leq z < 1.9$ \\
\hline
BL             & 11.0 [10.8,11.2] & 11.1 [10.8,11.6] & 11.8 [11.5,12.2] \\
NL            & 11.1 [10.8,11.1] & 11.3 [11.1,11.4]      & -- \\
BLAbs        & --\tnote{a}      & --               & 11.9 [11.5,12.3] \\
NLUnabs     & 10.9 [10.6,11.0] & 11.2 [11.0,11.7] & -- \\
\hline
\end{tabular}
\begin{tablenotes}
\footnotesize
\item[a] Single-source bins; values not statistically meaningful.
\end{tablenotes}
\end{threeparttable}
\end{table}

\begin{table}[ht]
\centering
\begin{threeparttable}
\caption{Median star formation rates ($\log \mathrm{SFR}$\,$[M_\odot\,\mathrm{yr}^{-1}]$) with interquartile ranges (IQR) for AGN populations in different redshift bins.
}
\label{tab:host_sfr}
\setlength{\tabcolsep}{1pt}
\begin{tabular}{lccc}
\hline\hline
Population & $\rm z<0.4$ & $\rm 0.4 \leq z < 0.8$ & $\rm 0.8 \leq z < 1.9$ \\
\hline
BL             & 1.36 [1.11,1.62] & 2.21 [1.89,2.67] & 3.16 [2.82,3.50] \\
NL            & 1.03 [0.59,1.48] & 1.50 [1.47,1.88]      & -- \\
BLAbs        & --\tnote{a}      & --               & 3.12 [2.62,3.61] \\
NLUnabs     & 1.41 [0.96,1.58] & 2.14 [1.93,2.35] & -- \\
\hline
\end{tabular}
\begin{tablenotes}
\footnotesize
\item[a] Single-source bins; values not statistically meaningful.
\end{tablenotes}
\end{threeparttable}
\end{table}

\begin{table}[ht]
\centering
\begin{threeparttable}
\caption{Median Eddington ratios ($\log \lambda_{\rm Edd}$) with interquartile ranges (IQR) for AGN populations in different redshift bins.}
\label{tab:host_edd}
\setlength{\tabcolsep}{0.05pt}
\begin{tabular}{lccc}
\hline\hline
Population & $\rm z<0.4$ & $\rm 0.4 \leq z < 0.8$ & $\rm 0.8 \leq z < 1.9$ \\
\hline
BL             & -1.37 [-1.66,-1.09] & -0.86 [-0.98,-0.61] & -0.59 [-0.73,-0.34] \\
NL            & -1.62 [-2.03,-0.93] & -1.06 [-1.31,-0.78]         & -- \\
BLAbs        & --\tnote{a}         & --                  & -0.86 [-0.93,-0.60] \\
NLUnabs     & -0.99 [-1.22,-0.73] & -0.37 [-0.57,0.00]  & -- \\
\hline
\end{tabular}
\begin{tablenotes}
\footnotesize
\item[a] Single-source bins; values not statistically meaningful.
\end{tablenotes}
\end{threeparttable}
\end{table}

\section{Summary}
\label{sec_summary}

We investigated the relation between optical obscuration and X-ray absorption in a combined SDSS--XMM sample of 241 AGN at $\rm z<1.9$, drawn from the 4XMM-DR11 catalogue \citep{Webb2020} and the SDSS DR16Q quasar catalogue \citep{Wu2022}. Bayesian X-ray spectral fits were obtained within the XMM2Athena framework \citep{Webb2023,Viitanen2025}, and host-galaxy properties were derived via SED fitting with \textsc{CIGALE} \citep{Boquien2019,Mountrichas2024d}. Our main results are:

\begin{itemize}

    \item [] Broad-line vs. narrow-line AGN:  
    Among the 241 AGN, 172 ($\sim$71\%) are classified as broad-line (BL) and 69 ($\sim$29\%) as narrow-line (NL). The majority of BL AGN ($\sim$94\%) show no signs of X-ray absorption, whereas $\sim$84\% of NL AGN (58/69) appear unabsorbed in X-rays (Fig. \ref{fig_fwhm_nh}), though this may partly reflect selection effects that disfavor strongly absorbed spectra.

    \item {Outlier populations:}  
    Two distinct subsets deviate from the standard optical–X-ray correspondence:  
    \begin{enumerate}
        \item BLAbs: Eleven BL AGN ($\sim$6\% of BL) exhibit significant X-ray absorption. Their SEDs remain type~1–like, indicating that the absorption arises from dust-free or geometrically complex gas rather than the classical torus.  
        \item NLUnabs: Fifty-eight NL AGN ($\sim$84\% of the NL sample) show no detectable X-ray absorption. Nearly half display type~1–like SEDs and high Eddington ratios, consistent with broad-line dilution by dust or host light rather than the intrinsic absence of a BLR.
    \end{enumerate}

    \item {Gas-to-dust ratios:}  
    At $\rm z<0.4$, NL AGN exhibit elevated $N_{\rm H}/E(B-V)$ ratios relative to BL AGN (median $\log N_{\rm H}/E(B-V)_{\rm AGN}=22.8$ vs.\ 21.1), consistent with their higher gas columns. BLAbs at $\rm z>0.8$ show extreme gas-to-dust ratios, revealing substantial X-ray–absorbing gas with little associated dust. NLUnabs have the lowest ratios, reflecting low $N_{\rm H}$ despite comparable dust extinction. A subset of the high $N_{\rm H}/E(B\!-\!V)$ outliers may trace late-stage mergers and potentially host dual AGN, although confirming this will require high-resolution X-ray/IR imaging or IFU spectroscopy.

    \item {Optical line diagnostics:}  
    At $\rm z<0.4$, NL AGN display higher Balmer decrements than NLUnabs, indicating stronger dust reddening. For [O\,III]/H$\beta$, only two BLAbs have measurements, preventing firm conclusions. Overall, optical diagnostics support enhanced reddening in NL AGN. 

    \item {Host galaxies and accretion:}  
    NL AGN tend to inhabit slightly more massive, lower-SFR hosts and show lower Eddington ratios than BL AGN, consistent with inefficient fueling and reduced cold-gas content. NLUnabs share BL-like SFRs and exhibit substantially higher Eddington ratios (up to $\log \lambda_{\rm Edd}\sim-0.4$), identifying them as efficiently accreting, dust-dominated systems. BLAbs match the BL population in SFR and accretion properties, suggesting that their excess X-ray absorption originates from external or complex absorbers rather than fueling differences.

\end{itemize}

These complementary diagnostics demonstrate that obscuration in AGN cannot be explained by orientation alone. Instead, it reflects a combination of circumnuclear geometry, host-scale dust, and temporal variability. The presence of BLAbs and NLUnabs highlights the need for multi-wavelength diagnostics to uncover the full diversity of AGN obscuration. Ongoing and upcoming wide-area surveys such as \emph{Euclid} and LSST will benefit from this integrated approach, combining X-ray spectroscopy, optical line diagnostics, and SED fitting to disentangle nuclear and host-scale contributions to AGN obscuration.

\begin{acknowledgements}
GM and FJC acknowledges funding from grant PID2021-122955OB-C41 funded by MCIN/AEI/10.13039/501100011033 and by “ERDF/EU”. This work was partially supported by the European Union's Horizon 2020 Research and Innovation program under the Maria Sklodowska-Curie grant agreement (No. 754510).

\end{acknowledgements}

\bibliography{mybib}
\bibliographystyle{aa}

\end{document}